\documentclass[submission, Phys]{SciPost}
\usepackage[normalem]{ulem}

\usepackage{graphicx}
\usepackage{amsmath}
\usepackage{amssymb}
\usepackage{color}
\usepackage{hyperref}

\newcommand{\tr}{\mathop{\text{tr}}}

\newcommand{\ve}{}
\newcommand{\tu}{t}

\DeclareMathOperator{\re}{Re}
\DeclareMathOperator{\im}{Im}
\begin{document}
\begin{center}{\Large \textbf{
Universal dynamics of spatiotemporal entrainment with phase symmetry
}}\end{center}
\begin{center}
Jorge Palacio Mizrahi\textsuperscript{1},
Danny Zilberg\textsuperscript{1},
Omri Gat\textsuperscript{1*}
\end{center}

\begin{center}
{\bf 1} The Racah Institute of Physics, The Hebrew University of Jerusalem, Jerusalem 9190401, Israel
\\
* omrigat@mail.huji.ac.il
\end{center}

\section*{Abstract}\textbf{We study the entrainment of a localized pattern by an external signal via its coupling to zero modes associated with broken symmetries. We show that when internal symmetries are broken, entrainment is governed by a multi-degree of freedom locking dynamical system that has a universal structure  defined by the internal symmetry group and its breaking. We derive explicitly the universal locking dynamics for entrainment of patterns breaking internal phase symmetry, and calculate the locking domains and the entrainment structure for the example of complex-Ginzburg-Landau solitons.}
\section{Introduction}
Entrainment is a central paradigm of nonlinear science, which can arise whenever a continuous symmetry is broken in two or more coupled systems. When the broken symmetry is time-translation invariance, entrainment leads to \emph{synchronization}, of which the simplest example is the frequency locking of limit cycles of nonlinear oscillators, either by external injection, as in the van der Pol circuit \cite{vanderpol}, or by mutual interaction, like Huygens' pendulum clocks \cite{pikovsky}, that can give rise to {collective} behavior when the number of synchronized oscillators is large \cite{kuramoto,kurarev,networks}. Synchronization occurs on more complex attractors as well, including quasiperiodic \cite{anis}, and chaotic \cite{chaos,pralitz,phasesync,hramov} attractors. Synchronization may take place in spatially extended system \cite{lin1,lin2}, of which recently studied examples  include dissipative optical soliton breathers \cite{breathers}, Kerr \cite{jang1,jang2} and laser \cite{lfc} frequency combs, and optomechanical \cite{phonon,opme} oscillators.

Entrainment of steady states breaking \emph{space}-translation invariance leads to wave-number locking, either by external injection \cite{lowe,coullet,manor1,manor}, or by mutual interaction \cite{miguez,mikhailov}. Patterns breaking both time- and space-translation invariance can be {spatiotemporally} entrained by spatially uniform  temporally periodic driving \cite{walgraef,rehberg,chate}, as well as by injection of traveling waves \cite{rudiger1,rudiger,beetz,rosen,kiel,gatkiel}.

The entrainment phenomenon is so ubiquitous because it is tied to spontaneous symmetry breaking. The entrainment dynamics is \emph{universal} for the same reason. For example, synchronization of a limit cycle always involves the locking dynamics of a phase, which is the variable associated with the zero mode (sometimes called Goldstone mode) that appears in the stability spectrum of the limit cycle as a result of the breaking of a continuous time-translation invariance to a discrete one; this mechanism is independent of the details of the dynamical system, and therefore the phase evolution is captured by the universal Adler equation \cite{pikovsky}.

Here we study the entrainment effect for localized patterns that break \emph{internal symmetries} in addition to space-translation invariance. Then, the stability spectrum of the pattern has internal-symmetry zero modes as well as a translation zero mode. A general perturbation couples to all of the modes, producing a multi-degree of freedom locking dynamical system that nevertheless has a universal structure for each set of broken internal symmetries.

We present two sets of results: We first derive the universal form of the locking dynamical system governing the entrainment of localized patterns with an internal phase symmetry, for which there are two symmetry generators, one for translations and one for phase shifts. It follows that entrainment in this class is governed by a two-degree of freedom dynamical system, which depends on two disparity parameters---relative velocity and frequency detuning. We show that for a given external injection there is a bounded locking domain in the space of possible disparities where the pattern can be stably entrained, and that the locking domain depends linearly on the strength of the perturbation.

Our second set of results is derived for the concrete example of pulse solutions of the cubic-quintic complex Ginzburg-Landau (QCGL) equation. We present locking diagrams showing the locking domains for several parameter choices of the QCGL equation. The diagrams exhibit the full complexity of a two-parameter family of two-degree of freedom dynamical systems, including multistability, and curves of saddle-node and Hopf bifurcations, tangent at Bogdanov-Taken bifurcation points.

\section{Locking dynamics with an internal phase symmetry.}
Consider a pattern-forming dynamical system in one space dimension
\begin{equation}\label{eq:dpsidtre}
\frac{\partial\Psi_u}{\partial t}=\mathcal{N}[\Psi_u]\ ,
\end{equation}
for the multicomponent field $\Psi_u(x)$.
The nonlinear operator $\mathcal{N}$ respects space-translation symmetry,
\begin{equation}\label{eq:trsym}
\mathcal{N}[\Psi(\cdot-\xi)](x)=\mathcal{N}[\Psi(\cdot)](x-\xi)
\end{equation}
for any shift $\xi$, and an internal symmetry
\begin{equation}\label{eq:trsym}
\mathcal{N}[G\Psi]=G\mathcal{N}[\Psi]
\end{equation}
for a continuous group of global operations $G$ that mix the components of $\Psi$ but are independent of $x$.

We assume that Eq.\ \eqref{eq:dpsidtre} has at least one stable spatially uniform, internal-symmetry invariant time-independent solution, and a stable pattern solution, $\Psi_0(x)$, that is localized near $x=0$, and tends to uniform solutions as $x\to\pm\infty$. In the frame where $\Psi_0(x)$ is stationary, $\mathcal{N}[\Psi_0]=0$. $\Psi_0$ breaks translation invariance by being localized, and we assume that it is not invariant under all $G$ operation, so that it breaks internal symmetry as well.

We next let the pattern interact with a weak external signal $F$ that is localized near $x=0$ at $t=0$, moving with a constant velocity $\ve v$, and undergoing internal symmetry transformations generated by $J$ at a  constant rate, relative to the free pattern, so that the equation of motion for the perturbed field $\Psi$ becomes
\begin{equation}\label{eq:dpsidtco}
\frac{\partial\Psi}{\partial t}=\mathcal{N}[\Psi]+\ve e^{-iJt}F(x-\ve v t)\ .
\end{equation}
When $t<0$ and $\ve |vt|$ is large, the only effect of the external injection is a slight perturbation in the tail of the pattern $\Psi_0$. However, when $t$ approaches zero, the injected signal overlaps with the non-trivial region of the localized pattern, and its interaction with the position and the internal symmetry of the pattern may lead to two distinct outcomes as $t\to+\infty$. If the interaction is weak, then the injected signal eventually overtakes the localized pattern with the interaction becoming negligible again when $|vt|$ is large, so that the net asymptotic result of the interaction is a  displacement $y_\infty^u$ and an internal symmetry transformation $G_\infty^u$ of the pattern:
\begin{equation}
\textbf{Unlocked:}\qquad\Psi(x,t)\mathop{\sim}_{t\to\infty}G_\infty^u\Psi_0(x-y_\infty^u)
\ .
\end{equation}

On the other hand, if the interaction is strong enough and effective, then the pattern becomes {entrained}: that is, as $t\to\infty$ it becomes stationary in the \emph{injection} frame, so that the pattern-signal interaction remains strong enough to maintain locking indefinitely with displacement $y_\infty^l$ and internal symmetry transformation $G_\infty^l$ {relative} to the entraining signal,
\begin{equation}\label{eq:locked-r}
\textbf{Locked:}\qquad\Psi(x,t)\mathop{\sim}_{t\to\infty}G_\infty^le^{J t}\Psi_0(x-\ve v t-y_\infty^l)
\ ,
\end{equation}

From this point we consider the special case of a pattern with an internal phase symmetry. Here $\Psi=(\psi,\psi^*)$, $\psi$ a complex-valued function, $\mathcal{N}[\Psi]=(\mathcal{O}[\Psi],\mathcal{O}[\Psi]^*)$, $\mathcal{O}$ a complex-function-valued differential operator, and ${}^*$ is complex conjugation. The symmetry operation is $G_\varphi\Psi=(e^{i\varphi}\psi,e^{-i\varphi}\psi^*)$, and the injection takes the form $e^{-iJt}F=(fe^{-i\omega t},f^*e^{i\omega t})$, where $f$ is a complex-valued function localized near $x=0$, and $\omega$ is the relative injection-pattern frequency.

The variables associated with the broken translation and phase symmetries are the position $\xi$ and phase $\varphi$ of the localized pattern $\psi_0$. Our next goal is derive the locking dynamical system for $\xi$ and $\varphi$. For this purpose note that even though $f$ is assumed small, $\xi$ and $\varphi$  may become large for $t\gg1$ since they are associated with broken symmetries, and therefore experience no restoring force. We accordingly assume that
\begin{equation}\label{eq:psi01}
\psi(x,t)=\Bigl(\psi_0 \bigl(x-\xi(\ve t)\bigr)+ \ve\psi_1 \bigl(x-\xi(\ve t),t \bigr)\Bigr)e^{-i\varphi(\ve t)}\ ,
\end{equation}
where $\psi_1$ is a small shape perturbation


Using this ansatz in Eq.\ \eqref{eq:dpsidtco} gives
\begin{equation}\label{eq:dPsi1dt}
\frac{\partial\Psi_1}{\partial t}+\frac{d\xi}{d\tu}\Psi_{\xi}+\frac{d\varphi}{d\tu}\Psi_{\phi}=\mathcal{L}\Psi_1+F\ ,
\end{equation}
where $\mathcal{L}$ is the linear stability operator of the pattern, i.e.\ the functional derivative of $\mathcal{N}$ with respect to $\Psi$, evaluated at $\Psi_0$, and $\Psi_{\xi}=-(\psi_0',(\psi_0')^*)$, $\Psi_{\varphi}=(-i\psi_0,i\psi_0^*)$, are the zero modes associated with the broken translation and phase symmetries  (respectively), $\mathcal{L}\Psi_{\xi}=\mathcal{L}\Psi_\varphi=0$.
The assumption that $\Psi_0$ is a stable pattern implies that the rest of the spectrum of $\mathcal{L}$ is in the left half of the complex plane. We make the stronger assumption that the spectrum of $\mathcal{L}$ other than the two zero eigenvalues is separated from the imaginary axis by a gap, so that $\Psi_1$ remains small.

The equations of motion for $\xi$ and $\varphi$ should be derived from \eqref{eq:dPsi1dt}; however, these variables are not fully defined by \eqref{eq:psi01}, because a small change in $\xi$ and $\varphi$ can be absorbed into $\psi_1$. This ambiguity is removed by requiring that
\begin{equation}
\langle\bar\Psi_{\xi},\Psi_1\rangle=\langle\bar\Psi_\varphi,\Psi_1\rangle=0\ ,
\end{equation}
where $\bar\Psi_{\xi}=(\bar\psi_{\xi},\bar\psi_{\xi}^*)$ and $\bar\Psi_\varphi=(i\bar\psi_\varphi,-i\bar\psi_\varphi^*)$ are the zero-eigenfunctions of $\mathcal{L}^\dag$ (the adjoint of $\mathcal{L}$) corresponding to $\Psi_\xi$ and $\Psi_\phi$, normalized with $\langle \bar\Psi_{\xi},\Psi_{\xi}\rangle=\langle \bar\Psi_\varphi,\Psi_\varphi\rangle=1$.

We can now project Eq.\ \eqref{eq:dPsi1dt} on $\bar\Psi_{\xi}$ and $\bar\Psi_\varphi$, obtaining
\begin{align}
\frac{d\xi}{d\tu}&=\langle \bar\Psi_{\xi},F\rangle\equiv c_{\xi}(\xi-v\tu,\varphi-\omega\tu)\ ,\label{eq:2dsystema}\\\frac{d\varphi}{d\tu}&=\langle \bar\Psi_\varphi,F\rangle\equiv c_\varphi(\xi-v\tu,\varphi-\omega\tu)\ ,\label{eq:2dsystemb}
\end{align}
and change variables to the position and phase shifts, $y=\xi-v\tu$ and $\theta=\varphi-\omega\tu$ (respectively), finally obtaining
\begin{align}
\frac{dy}{d\tu}&= c_{\xi}(y,\theta)-v\label{eq:dydtau}\ ,\\
\frac{d\theta}{d\tu}&= c_\varphi(y,\theta)-\omega\ ,\label{eq:dthetadtau}
\end{align}
where the real-valued functions $c_\xi$ and $c_\varphi$ are localized in $y$ and periodic in $\theta$.
Equations (\ref{eq:dydtau},\ref{eq:dthetadtau}) comprise the universal locking dynamical system for localized patterns with an internal phase symmetry, perturbed by an external signal. (Stable) fixed points of the locking system correspond to (stable) entrained steady states of the pattern. The examples studied below show that the structure and properties of the stable entrained states can be quite varied and complex. Nevertheless, some general features follow directly from the equations.

First note that the functions $c_{\xi},c_\varphi$ are bounded: $v_{\min}\le c_{\xi}\le v_{\max}$, $\omega_{\min}\le c_\varphi\le \omega_{\max}$. Hence, entrainment is  only possible inside a bounded domain in the $\omega$-$v$ parameter plane.

To go beyond this simple observation it is convenient to use the inner product
\begin{equation}
\langle \Phi,\Psi\rangle\equiv\langle (\phi,\tilde\phi),(\psi,\tilde\psi)\rangle=\frac{1}{2}\int_{-\infty}^\infty dx(\phi^*\psi+\tilde\phi^*\tilde\psi)\ ,
\end{equation}
so that
\begin{align}
c_{\xi}(y,\theta)&=\re\mathcal{I}\equiv g(y)\cos\left[\theta-\alpha(y)\right]\ ,\label{eq:cx}\\
c_{\varphi}(y,\theta)&=-\im\mathcal{I}\equiv h(y)\cos\left[\theta-\beta(y)\right]\ ,\label{eq:cphi}
\end{align}
where
\begin{equation}
\mathcal{I}=e^{i\theta}\int_{-\infty}^{\infty}\bar\psi_{\xi,\varphi}(x)^*f(x+y)dx\ .
\end{equation}
Entrained steady states are solutions $(\hat y,\hat\theta)$ of the system
\begin{equation}
 h(\hat y)\cos\bigl(\hat \theta-\alpha(\hat y)\bigr)=v\label{eq:dydtau=0}\ ,\quad
g(\hat y)\cos\bigl(\hat \theta-\beta(\hat y)\bigr)=\omega\ ;
\end{equation}
eliminating $\hat\theta$ gives the equation
\begin{equation}\label{eq:vomega}
v=(h/g)(\omega\cos\gamma\pm\sin\gamma\sqrt{g^2-\omega^2}\ )\ ,
\end{equation}
where $\gamma=\beta-\alpha$, and all functions are evaluated at $\hat y$.

Choices of disparity parameters $(\omega,v)$ for which \eqref{eq:vomega} has one or more solutions constitute the \emph{locking domain}, where entrainment is possible for a given injection signal $f$.
Clearly $|\omega|<\max(|g|)$ inside the locking domain. For any $\omega$ in this interval, \eqref{eq:vomega} defines a function $v_\omega(y)$ defined for all $y$ such that $g^2(y)>\omega^2$, whose range is a collection of one or more intervals, and the locking domain can be pieced together from these intervals by running over the allowed values of $\omega$; several examples are shown below.

The \emph{stable} locking domain consists of those disparity combinations $(\omega,v)$ in the locking domain for which the eigenvalues of the Jacobian matrix
\begin{equation}\label{eq:J}
J=\begin{pmatrix}\partial_yc_{\xi}&\partial_{\theta}c_{\xi}\\\partial_yc_{\phi}&\partial_{\theta}c_{\phi}\end{pmatrix}
\end{equation}
have negative real parts for at least one entrained steady state. For such entrained states $\tr J<0$ and $\det J>0$, and therefore at the boundary of the stable locking region either $\tr J=0$ or $\det J=0$. The latter equality, together with the fixed-point condition (\ref{eq:dydtau=0}), implicitly defines a curve or curves of saddle-node bifurcations on the stability boundary, and the former defines  a curve or curves of Hopf bifurcations; the curves of both kinds meet at points of Bogdanov-Takens bifurcations where both eigenvalues are zero \cite{arnold}.
Note that the locking functions $g$ and $h$, as well as the Jacobian matrix $J$ depend linearly on $f$, so that the locking domains and stable locking domains scale linearly with the injection amplitude.
\begin{figure}[tb]
\centering
{\includegraphics[scale=0.25]{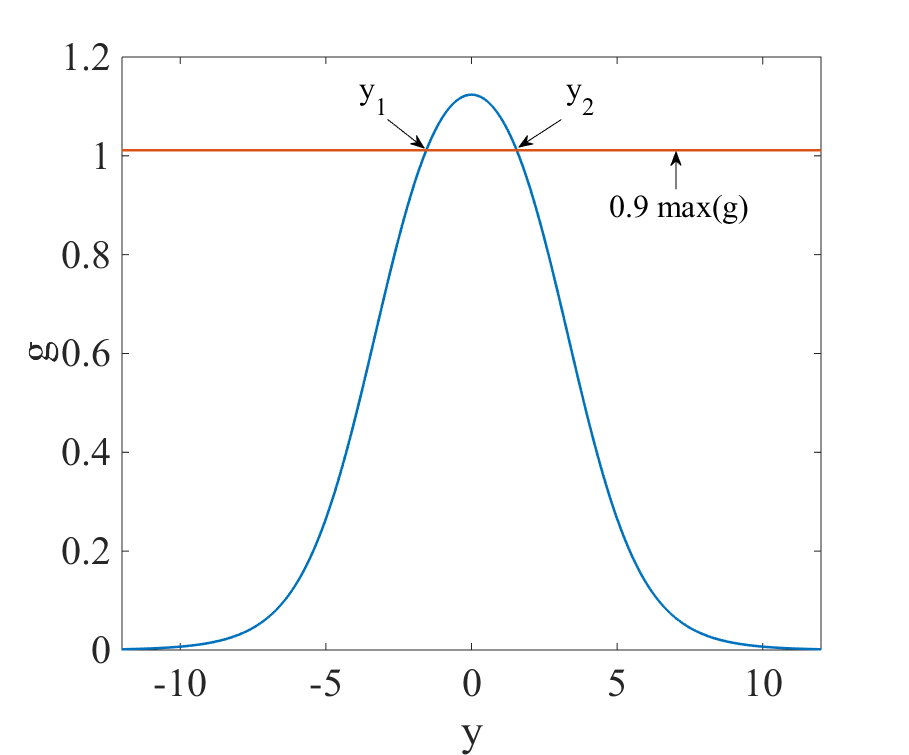}\quad
\includegraphics[scale=0.25]{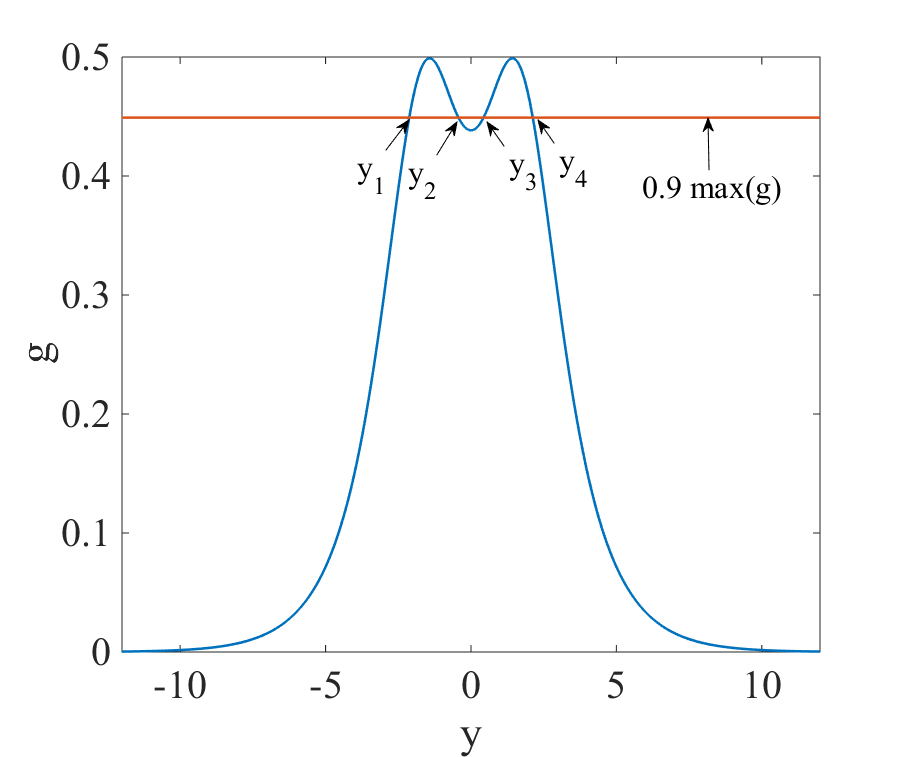}}
\caption{\small The locking function $g$ defined by \eqref{eq:cx} for a steady state QCGL pulse entrained by an externally injected Gaussian pulse. The $g$ value of the red line represents an admissible value of frequency disparity $\omega$, and the range of $y$ for which $g^2(y)>\omega^2$ represent possible entrained steady states of the system for this value of $\omega$. When the injected signal is broad (left), $g$ is unimodal, and there is a single interval of locked states for each $\omega$, and when the injected signal is narrow (right), $g$ is bimodal, and are two disjoint intervals of locked states for sufficiently large $\omega$. The parameters of the QCGL pulse are $\beta=0.8$, $\delta=-0.1$, $\epsilon=0.8$, $\mu=-0.5$, $\nu=-0.1$, $\omega_0=0.638$. (see equation \ref{eq:qcgl} for the parameter definitions); Gaussian injection parameters are $a=0.1$ (here and below) and $c=0.2$ (left) and $c=1.3$ (right).\label{fig:gvsy}}
\end{figure}
\begin{figure}[tb]
\centering
{\includegraphics[scale=0.25]{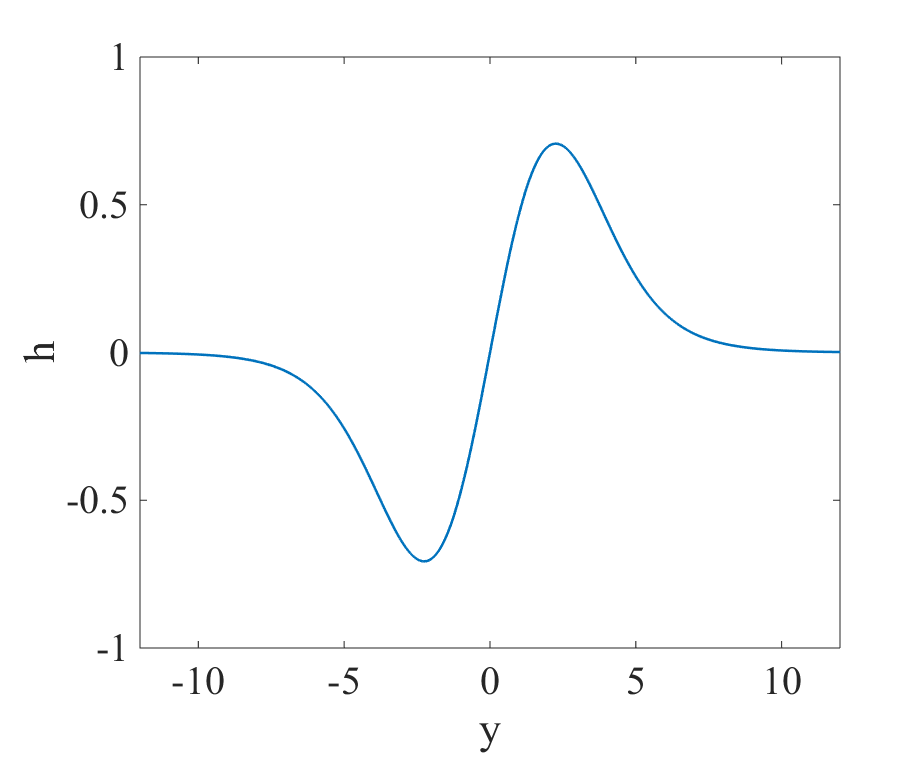}\quad
\includegraphics[scale=0.25]{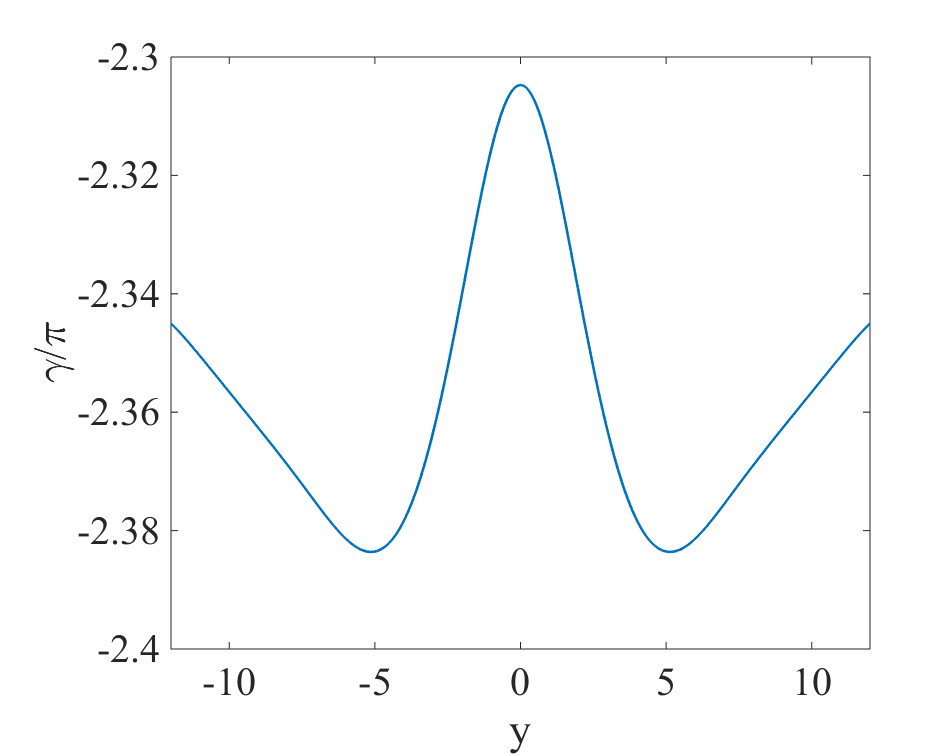}}
\caption{\small The locking functions $h$ (defined in equation \ref{eq:cphi}) and $\gamma=\alpha-\beta$ (equation \ref{eq:cx}, \ref{eq:cphi}), for entrainment steady state QCGL pulse by a Gaussian. The QCGL and injection parameters are as in figure \ref{fig:gvsy}.\label{fig:hgvsy}}
\end{figure}

\section{Entrainment of complex Ginzurg-Landau pulses} The cubic-quintic complex Ginzburg-Landau (QCGL) equation, defined by the nonlinear operator
\begin{equation}\label{eq:qcgl}
\mathcal{O}[\Psi]=(\delta-i\omega_0)\psi+(\beta+{i}/{2}){\partial^2\psi}/{\partial x^2}+(\epsilon+i)|\psi|^2\psi+\left(\mu+i\nu\right)|\psi|^4\psi\ ,
\end{equation}
with real parameters $\delta$, $\beta$, $\epsilon$ $\mu<0$, and $\nu$, has internal phase symmetry, and is known to have stable pulse solutions in some parameter ranges \cite{soto}, which become stationary for appropriately chosen $\omega_0$.
\begin{figure}[tb]
\centering
{\includegraphics[scale=0.25]{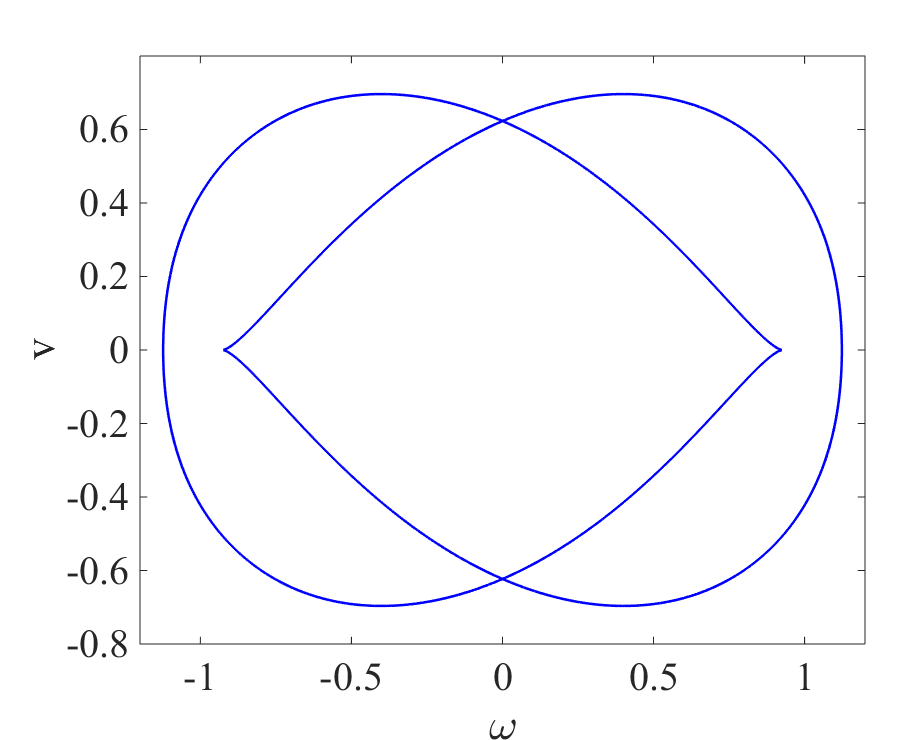}\quad
\includegraphics[scale=0.25]{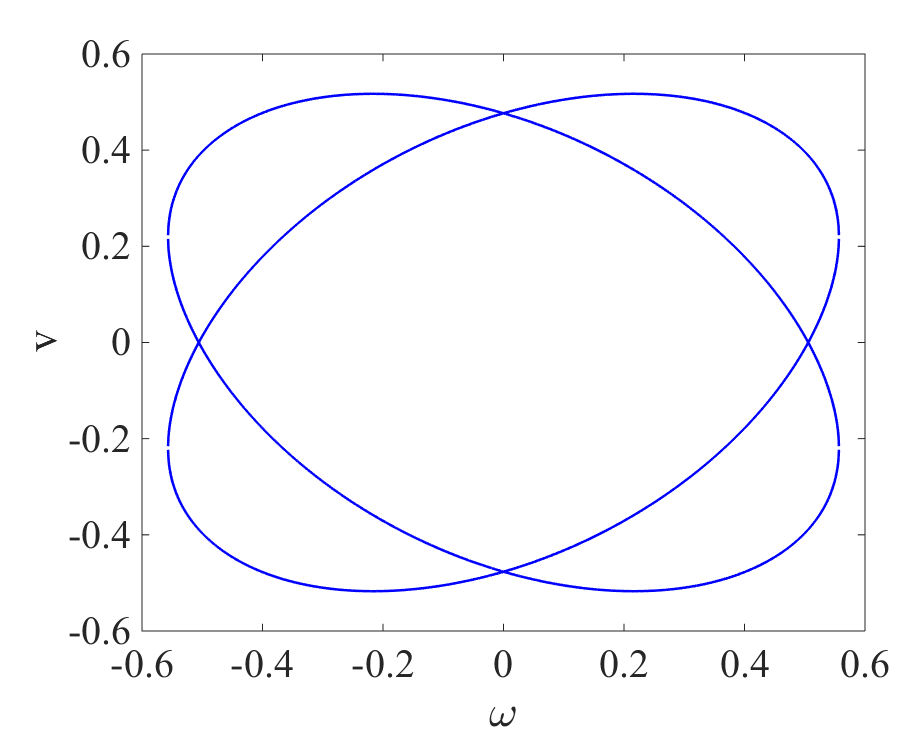}}
\caption{\small Locking diagrams showing boundaries of branches of steady states of QCGL pulses entrained by externally injected Gaussian pulses, depending smoothly on the frequency disparity $\omega$ and velocity disparity $v$. QCGL parameters are $\beta=0.8$, $\delta=-0.1$, $\epsilon=0.8$, $\mu=-0.5$, $\nu=-0.1$, $\omega_0=0.638$. (see equation \ref{eq:qcgl} for the parameter definitions); Gaussian injection parameters are $a=0.1$ and $c=0.2$ (left) and $c=1$ (right).\label{fig:lobes}}
\end{figure}
We will consider entrainment of QCGL pulses by a Gaussian pulse $f(x)=ae^{-cx^2}$, $a ,c>0$. Thanks to the parity symmetry of the QCGL equation, its pulse solutions are even, $\psi_0(-x)=\psi_0(x)$, making $h$ an odd function and $g$, $\alpha$, and $\beta$ even functions of $y$. These symmetries imply that if $(\hat y,\hat\theta)$ is a fixed point of (\ref{eq:dydtau}--\ref{eq:dthetadtau}) with detuning parameters $v,\omega$ then so are $(-\hat y,\hat\theta)$ with $-v,\omega$, and $(-\hat y,\hat\theta+\pi)$ with $v,-\omega$, and therefore the locking domain is symmetric with respect to the sign flips of both $v$ and $\omega$. Furthermore, it follows from \eqref{eq:J} that under a sign flip of $\hat y$ the (off-)diagonal elements of $J$ remain invariant (change sign), so that $\tr J$ and $\det J$ are unchanged, while a $\pi$ shift of $\hat\theta$ makes all the elements of $J$ flip sign. Consequently, the \emph{stable} locking domain is invariant under a sign flip of $v$, while at most one of the locking states related by a sign change of $\omega$ can be stable.

Figures \ref{fig:gvsy} and \ref{fig:hgvsy} present numerical examples of the locking functions. As shown in figure \ref{fig:gvsy}, the even function $g$ can be either unimodal with a maximum at $y=0$, or bimodal, with a minimum at $y=0$. In the former case, there is a single interval of velocity locking intervals $v$ for each admissible $\omega$, determined by \eqref{eq:vomega}; in the latter case, there are two locking intervals for each $\omega$ larger than the local minimum of $g$ at $y=0$. Consequently, the locking domains for broad and narrow injections are quite different, as shown below.
Unlike $g$, the locking functions $h$ and $\gamma$ have the same qualitative structure for all QCGL and injection parameters that we tested; examples are shown in figure \ref{fig:hgvsy}.

\begin{figure}[tb]
\centering
\includegraphics[scale=0.25]{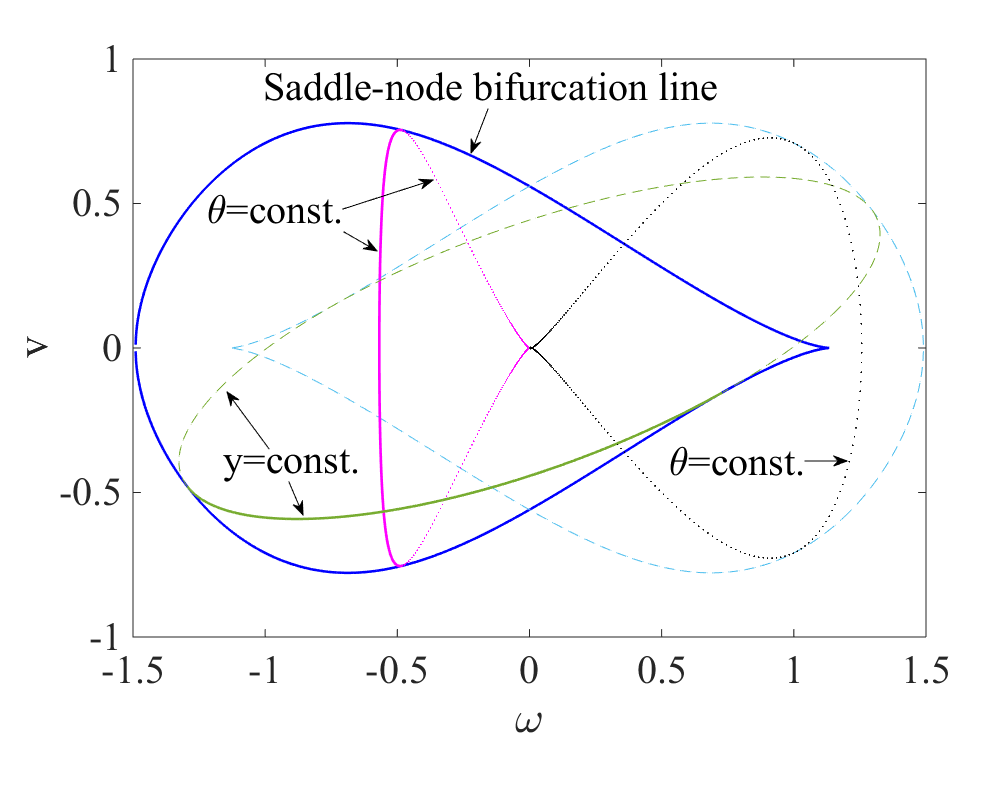}\quad
\includegraphics[scale=0.25]{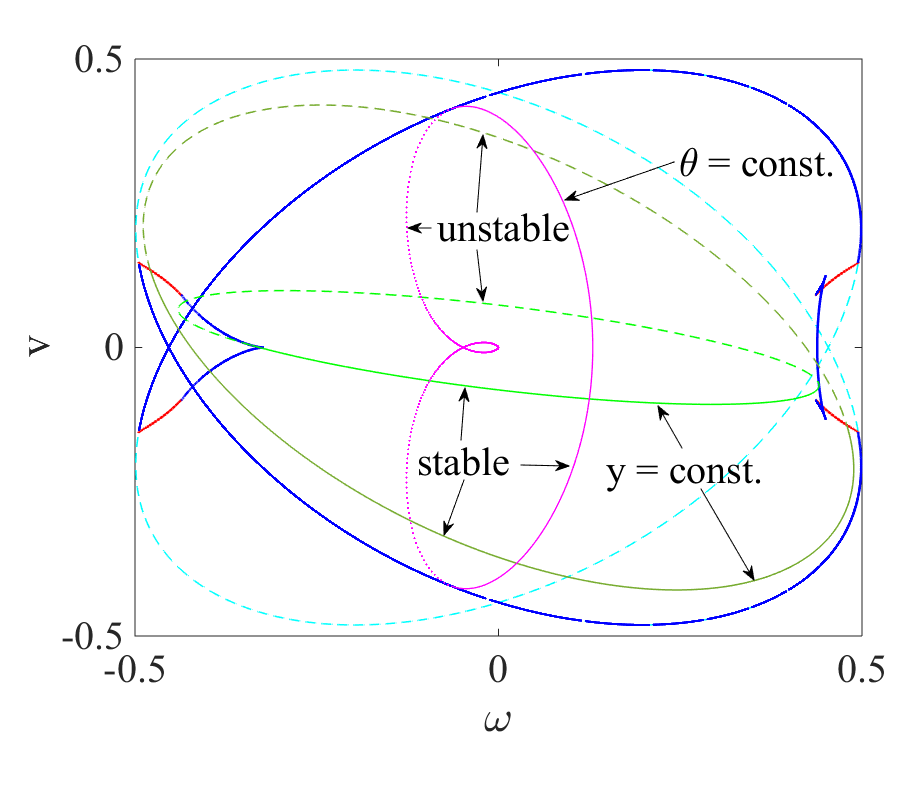}
\caption{{\small Locking diagrams for QCGL and injection parameters $\beta=0.5$, $\delta=-0.1$, $\epsilon=1.4$, $\mu=-0.1$, $\nu=0.2$, $\omega_0=0.638$, $c=3$ (left), and $\beta=0.8$ $\delta=-0.1$, $\epsilon=0.8$, $\mu=-0.5$, $\nu=-0.1$, $c=1.3$ (right), in the plane of disparity parameters $\omega$ and $v$. The boundaries of the stable part of the principal branch of entrained steady states consist of saddle-node bifurcations (solid blue curves), and Hopf bifurcations (solid red curves in the right panel). The two families of bifurcation meet Bogdanov-Takens bifurcation points. Unstable branch boundaries are shown as dashed light blue curves. Green (pink) interior curves show sub-branches of entrained states with constant injection-pulse displacement $\hat y$ (phase shift $\hat\theta$) respectively; solid (dashed/dotted) parts of the curves show stable (unstable) parts, respectively of the  sub-branches. \label{fig:bifurcation2}}}
\end{figure}

\begin{figure}[tb]
\centering
\includegraphics[scale=0.25]{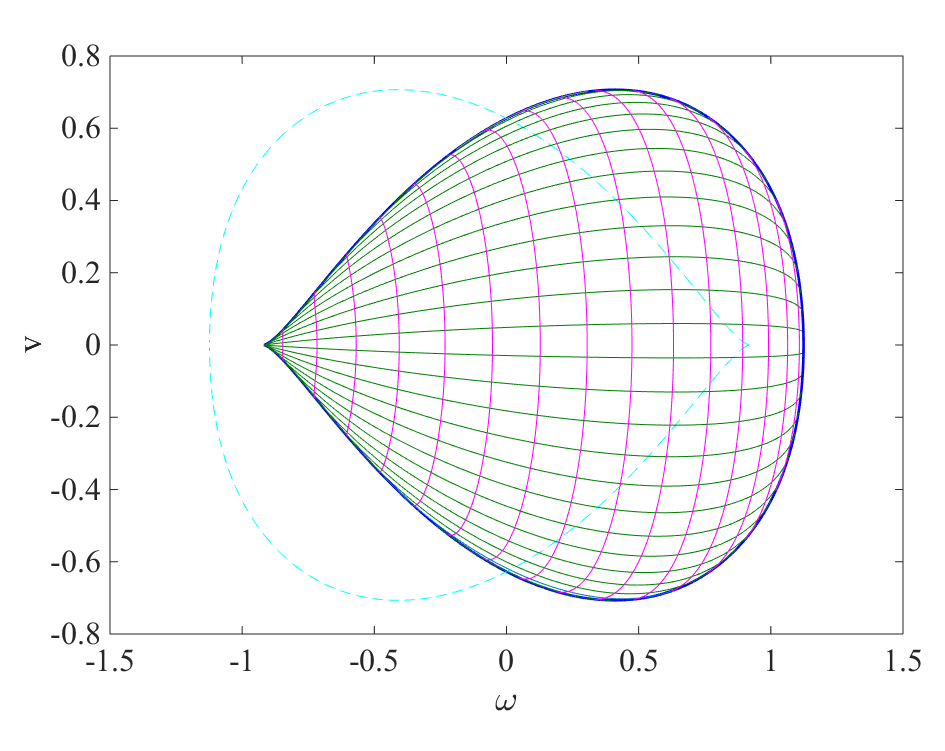}\quad\includegraphics[scale=0.23]{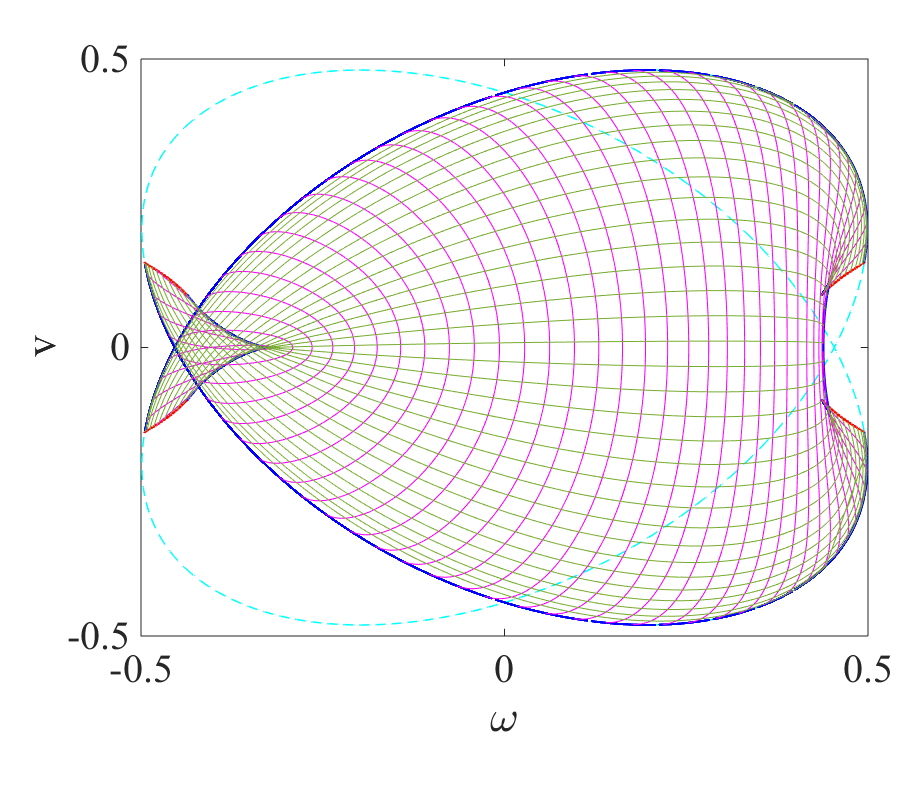}
\caption{{\small Locking diagrams for QCGL and injection parameters  $\beta=0.8$, $\delta=-0.1$, $\epsilon=0.8$, $\mu=-0.5$, $\nu=-0.1$, $\omega_0=0.638$, and $c=0.2$ (left) and $c=1.3$ (right), in the plane of disparity parameters $\omega$ and $v$. The stable parts of the principal branches are coordinatized by the green curves of constant-$\hat y$ entrained states and by the pink curves of constant-$\hat\theta$ states. Branch and stability boundaries are marked as in figure \ref{fig:bifurcation2}. \label{fig:pulse2}}}
\end{figure}

Figure \ref{fig:lobes} shows two examples of locking domains of the same QCGL pulse entrained by injection of Gaussians of different widths; the blue curves are boundaries of branches of entrained steady states that are solutions of (\ref{eq:dydtau=0}), consisting of saddle-node bifurcation points. In the left panel the injected pulse is wide, making the locking function $g$ unimodal, with a single interval of solutions of \eqref{eq:vomega} for each $\omega$, as in the example shown in the left panel of figure \ref{fig:gvsy}; it gives two overlapping teardrop-shaped locking branches that are individually invariant under $v\leftrightarrow-v$, and are mapped to each other by $\omega\leftrightarrow-\omega$. Of the two branches, exactly one, the \emph{principal branch}, consists entirely of stable locked states. The similarly structured locking diagram in the left panel of Fig.\ \ref{fig:bifurcation2} shows the boundary of the principal branch in solid blue curve as in figure \ref{fig:lobes}, and the boundary of the secondary, unstable, branch is marked with a dashed blue curve. This figure also shows contours of constant-$\hat y$ and $\hat\theta$ locked states in green and pink (respectively), with solid (dashed) curves representing stable (unstable) solutions respectively. Note that equations \eqref{eq:dydtau=0} imply that constant-$\hat y$ curves are ellipses or ellipse arcs. The left panel of Fig.\ \ref{fig:pulse2} shows the locking diagram of Fig.\ \ref{fig:lobes} (left) with the principal branch coordinated by $\hat y$ (green curves) and $\hat\theta$ (pink curves).

In contrast with these example, the right panels of figures \ref{fig:lobes}--\ref{fig:pulse2} show locking diagrams obtained for a narrow injected signal, which makes the locking function $g$ bimodal, and produces two disjoint $v$ locking intervals for large enough $\omega$, as in the example shown in the right panel of figure \ref{fig:gvsy}. Consequently, the two ovals in the diagram of the right panel of figure \ref{fig:lobes} are boundaries of a \emph{single} branch---the principal branch, that is invariant under sign changes of both $v$ or $\omega$, and is double-valued where the oval interiors overlap. However, in this case principal branch locking states undergo Hopf bifurcations, so only some of them are stable. The Hopf bifurcation curves that are marked in red in the locking diagram  of Fig.\ \ref{fig:bifurcation2} (right), make up the boundary of the stable locking domain together with the saddle-node bifurcation curves that are shown in solid blue, as before. The two kinds of bifurcation curves meet at Bogdanov-Takens bifurcation {points} \cite{arnold}. Figure \ref{fig:bifurcation2} (right) also shows constant-$\hat y$ and $\hat\theta$ entrained state curves in the same manner as they are shown in figure \ref{fig:bifurcation2} (left). The locking diagram of figure \ref{fig:pulse2} (right) shows the Hopf- and saddle-node-bifurcation parts of the boundary of the stable locking domain as in figure \ref{fig:bifurcation2} (right) and its $\hat y$-$\hat\theta$ coordination as in figure \ref{fig:pulse2} (left).

Finally we note that since the locking domains scale linearly with the amplitude of injection, the locking diagrams of Figs.\ \ref{fig:lobes}--\ref{fig:pulse2} can be viewed as two-dimensional section of \emph{cones} in a three-dimensional parameter space of $\omega$, $v$ and injection amplitude.


\section{Conclusions}
Entrainment takes place when nonlinear systems with broken continuous symmetries interact. The fundamental property of entrainment phenomena is the sharp locking transition of the symmetry-breaking variables. The locking transition is a manifestation of a bifurcation in the locking dynamical system that governs the evolution of the symmetry breaking variables. In standard entrainment scenarios a single symmetry is broken---most notably time-translation invariance in synchronization. However, patterns in extended systems often break internal symmetries as well as space-translation invariance. Here we have shown that when localized patterns with an internal symmetry are spatiotemporally forced, the locking dynamics becomes a multiple-degree of freedom system, whose form is universal within a symmetry class.

Focusing on the case of externally forced patterns with an internal phase symmetry, we found nonstandard locking behavior, including entrainment bistability, Hopf, and Bogdanov-Takens bifurcations. Even though here we studied in detail only the stationary locked states, it is clear that much of the complexity of two-degrees of freedom dynamical systems can be realized in the locking dynamics; for example, the occurrence of Hopf bifurcations implies that oscillatory ``breather'' entrainment can be realized on limit cycles of the locking system.

The shown examples of locking domains are two-dimensional sections of cones that are actually tips of three-dimensional Arnold tongues \cite{tongues}. In single-variable locking Arnold tongues emerge at all commensurate frequencies, yielding locking diagrams with a fractal structure. We conjecture that beyond the fundamental entrainment that was studied here,  entrainment of patterns with internal symmetries exhibits a complex structure of harmonic entrainment.

\paragraph{Funding information}
This work was supported by the Israel Science Foundation.

\end{document}